\documentclass[reprint,superscriptaddress,pra]{revtex4-1}
\usepackage{xcolor}
\usepackage{graphicx}
\usepackage{hyperref}
\usepackage{dcolumn}
\usepackage{makecell}
\usepackage[T1]{fontenc}
\newcolumntype{d}[1]{D{.}{.}{#1}}
\newcommand\mc[1]{\multicolumn{1}{c}{#1}}

\begin{document}

\title{Ultra-high finesse cavity-enhanced spectroscopy for accurate tests of quantum electrodynamics for molecules}

\author{M.~Zaborowski} \email{mzab@doktorant.umk.pl}
\author{M.~S\l{}owi\'{n}ski}
\author{K.~Stankiewicz}
\affiliation{Institute of Physics, Faculty of Physics, Astronomy and Informatics, Nicolaus Copernicus University in Toru\'{n}, Grudzi\k{a}dzka 5, 87-100 Toru\'{n}, Poland}

\author{F.~Thibault}
\affiliation{Univ Rennes, CNRS, IPR (Institut de Physique de Rennes) - UMR 6251, F-35000 Rennes, France}

\author{A.~Cygan}
\author{H.~J\'{o}\'{z}wiak}
\author{G.~Kowzan}
\author{P.~Mas\l{}owski}
\author{A.~Nishiyama}
\author{N.~Stolarczyk}
\author{S.~W\'{o}jtewicz}
\author{R.~Ciury\l{}o}
\author{D.~Lisak}
\author{P.~Wcis\l{}o} \email{piotr.wcislo@fizyka.umk.pl}
\affiliation{Institute of Physics, Faculty of Physics, Astronomy and Informatics, Nicolaus Copernicus University in Toru\'{n}, Grudzi\k{a}dzka 5, 87-100 Toru\'{n}, Poland}

\date{\today}

\begin{abstract}

We report the most accurate measurement of the position of the weak quadrupole S(2) 2--0 line in D$_2$.
The spectra were collected with a frequency-stabilized cavity ring-down spectrometer \mbox{(FS-CRDS)} with an ultra-high finesse optical cavity ($\mathcal{F}$~=~637~000) and operating in the frequency-agile, rapid scanning spectroscopy (FARS) mode.
Despite working in the Doppler-limited regime, we reached 40~kHz of statistical uncertainty and 161~kHz of absolute accuracy, achieving the highest accuracy for homonuclear isotopologues of molecular hydrogen.
The accuracy of our measurement corresponds to the fifth significant digit of the leading term in QED correction.
We observe 2.3$\sigma$ discrepancy with the recent theoretical value.

\end{abstract}

\maketitle

\section{\label{sec:level1}Introduction}
Molecular hydrogen, in the view of its simplicity, is well suited for testing quantum electrodynamics (QED) for molecules \cite{PhysRevA.100.032519,PhysRevA.98.052506} as well as for searching for new physics beyond the Standard Model such as new forces \cite{PhysRevD.87.112008} or extra dimensions \cite{Salumbides_2015}.
Furthermore, molecular hydrogen possesses a wide structure of ultra-narrow rovibrational transitions \cite{Wolniewicz_1998} with different sensitivities to the proton charge radius and proton-to-electron mass ratio.
Therefore, the recent large progress in both theoretical \cite{C8CP05493B,PhysRevA.100.032519,PhysRevA.98.052506} and experimental \cite{Diouf:19,PhysRevLett.120.153001,PhysRevA.98.022516,MONDELAIN20165,WCISLO201841} determinations of the rovibrational splitting in different isotopologues of molecular hydrogen makes it a promising system for adjusting several physical constants \cite{RevModPhys.88.035009,Beyer79}.
The most accurate measurements were performed for the HD isotopologue with absolute accuracy claimed to be 20~kHz \cite{PhysRevLett.120.153002} and 80~kHz \cite{PhysRevLett.120.153001} for the R(1) 2--0 line.
Such accuracy was obtained by saturating the transition and measuring the sub-Doppler structure.
These two results \cite{PhysRevLett.120.153002,PhysRevLett.120.153001} differ, however, by almost 1~MHz.
Recently, it was reported \cite{Diouf:19} that the uncertainty from Ref.~\cite{PhysRevLett.120.153002} was underestimated due to a complex hyperfine structure and should be 50~kHz.
Although HD possesses electric dipole transitions, they are extremely weak and a high-finesse optical cavity (with finesse of the order of 10$^5$) is necessary to build a sufficiently large intracavity power and saturate the transitions.
Homonuclear isotopologues (due to the symmetry of molecules) do not possess even weak electric dipole lines in the ground electronic state, and direct studies of the rovibrational structure were performed on the quadrupole transitions, which are almost 3 orders of magnitude weaker than the dipole lines observed in HD.
For this reason, their transitions were not saturated and measurements were performed with the cavity-enhanced Doppler-limited techniques.
The most accurate measurements for the D$_2$ isotopologue were performed for the first overtone (the S(2) line) and reached an absolute accuracy of 500~kHz \cite{MONDELAIN20165} and 400~kHz \cite{WCISLO201841}.
For the case of the H$_2$ isotopologue, the most accurate measurements that directly probe the rovibrational lines reached an accuracy of 6.6~MHz \cite{PhysRevA.93.022501} (performed for second overtone) and the measurement based on the subtraction of the energies of two electronic transitions provided the energy of the fundamental band lines with an accuracy of 4.5~MHz \cite{PhysRevLett.110.193601,NIU201444}.

In this letter, we report the most accurate measurement of the position of the weak quadrupole S(2) 2--0 line in D$_2$.
The spectra were collected with a frequency-stabilized cavity ring-down spectrometer \mbox{(FS-CRDS)} linked to an optical frequency comb (OFC) referenced to a primary frequency standard \cite{doi:10.1063/1.4952651,Cygan:19}.
We developed an ultra-high finesse optical cavity ($\mathcal{F}$~=~637~000) and we implemented the frequency-agile, rapid scanning spectroscopy (FARS) \cite{Truong2013FrequencyagileRS}, which allowed us to reduce the absorption noise from 3$\times$10$^{-10}$~cm$^{-1}$ in our previous study \cite{WCISLO201841} to 8$\times$10$^{-12}$~cm$^{-1}$ in the present work.
In spectra analysis, we used the speed-dependent billiard-ball profile (SDBBP) \cite{PhysRevA.65.012502}, whose parameters were determined in Ref.~\cite{WCISLO201841} based on the analysis that merged both \textit{ab~initio} calculations and high-pressure measurement.
This allowed us to reduce the systematic effects related to collisional perturbation to the shapes of molecular lines \cite{PhysRevA.93.022501}.
Despite operating in the Doppler-limited regime we reached 40~kHz of statistical uncertainty and 161~kHz of absolute accuracy of the S(2)~2--0 transition energy determination achieving the highest accuracy for homonuclear isotopologues of molecular hydrogen and only 3.2 times lower accuracy compared to the most accurate sub-Doppler measurements in HD \cite{Diouf:19} whose line intensities are almost 3 orders of magnitude stronger \cite{Diouf:19,C0CP00209G}.
The accuracy of our measurement corresponds to the fifth significant digit of the leading term in QED correction.
We observe 2.3$\sigma$ discrepancy between our experimental and most recent theoretical value \cite{PhysRevA.98.052506}.

\section{\label{sec:level1}Experimental Setup}
\begin{figure}[t]
\centering \includegraphics[width=0.49\textwidth]{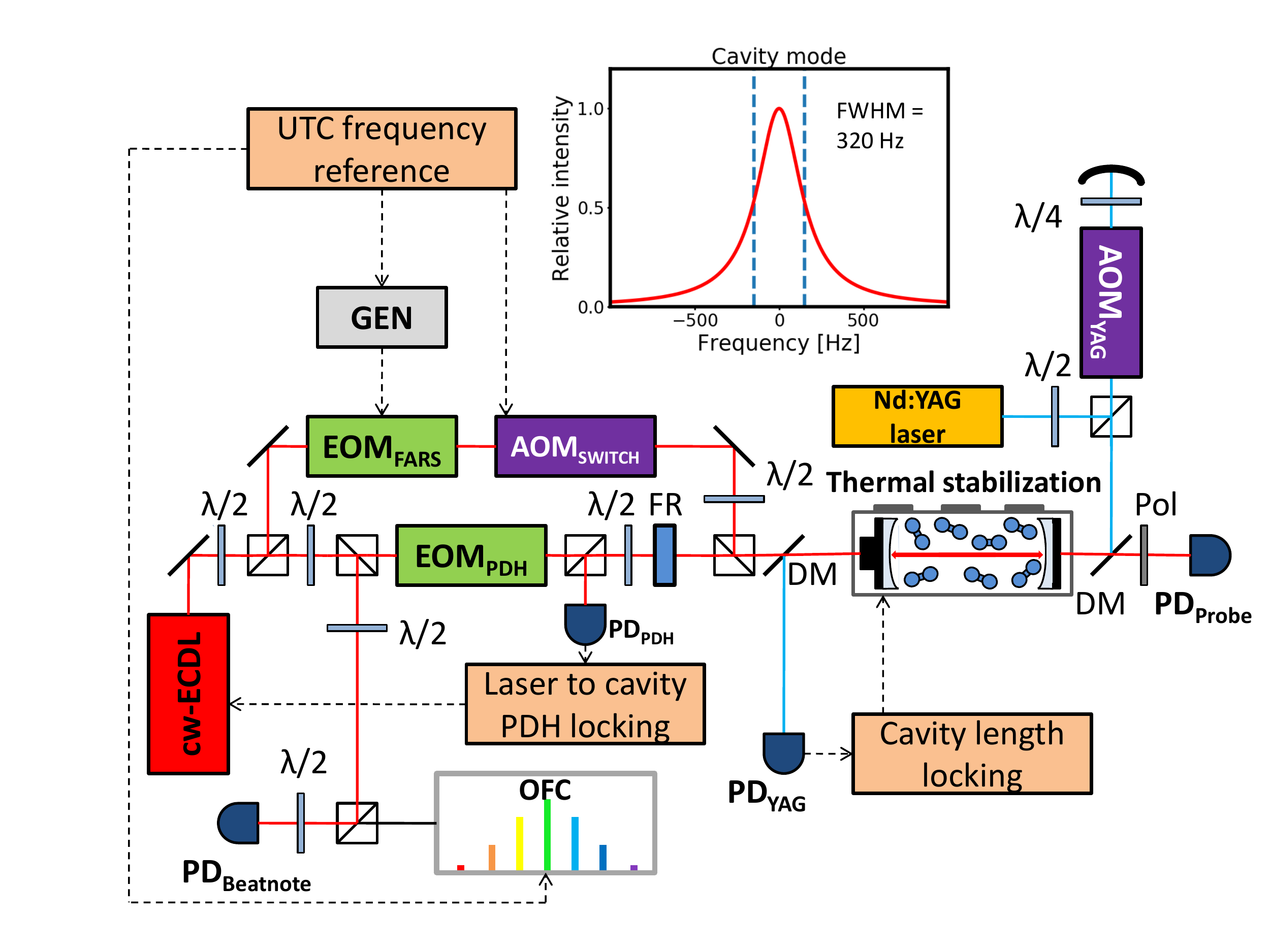}
\caption{The frequency-stabilized cavity ring-down spectrometer \mbox{FS-CRDS} referenced to the primary frequency standard UTC (AOS).
Light from a continuous-wave external cavity diode laser (cw-ECDL) is polarization split into two beams.
One of them is used for PDH locking of the ECDL frequency to the cavity mode and to determine its absolute frequency by measuring the heterodyne beat with the OFC.
EOM$_{PDH}$ modulates the phase of light to create the PDH error signal.
Second beam probes the gas sample inside the cavity by ring-down decay signals initiated by an acousto-optic modulator (AOM$_{SWITCH}$).
The cavity length is actively stabilized to the Nd:YAG laser.
The EOM$_{FARS}$, also referenced to the UTC (AOS), is used for fast full-spectrum scanning by stepping a laser sideband to successive optical cavity modes.
$FR$ -- Faraday rotator, $GEN$ -- generator, $DM$ -- dichroic mirror, $Pol$ -- polarizer, and $PD$ -- photodiodes.
}
\label{setup}
\end{figure}

The experimental setup is shown in Fig.~\ref{setup}.
The enhancement cavity of the \mbox{FS-CRDS} spectrometer is length-stabilized with respect to the I$_2$-stabilized Nd:YAG laser operating at 1064~nm to prevent the thermal drift of the cavity modes.
The deuterium sample has a purity of 99.96\%.
The length of the cavity is 73.5~cm which corresponds to a free spectral range (FSR) of 204~MHz.
We use an acousto-optic modulator (AOM$_{YAG}$) arranged in a double-pass configuration to control the cavity length and, hence, tune the laser frequency on a denser grid than the FSR spacing.
The laser is linked to an OFC which is referenced to the Coordinated Universal Time (UTC), the primary time standard provided by the Astro-Geodynamic Observatory in Borowiec (Poland) \cite{doi:10.1063/1.4952651}.

In this work, we improved the previous experimental setup \cite{WCISLO201841} by developing a cavity with much higher finesse and implementing the frequency-agile rapid scanning spectroscopy (FARS) technique \cite{Truong2013FrequencyagileRS}.
We improved the finesse of the previous cavity \cite{WCISLO201841} from $\mathcal{F}$~=~4$\times$10$^4$ to 6.4$\times$10$^5$, hence the light-molecule interaction path increased by more than one order of magnitude.
It corresponds to an ultra-narrow mode width of 320~Hz (0.5~ms \mbox{ring-down} decay time) and, to our knowledge, it is the highest finesse used in molecular spectroscopy measurements.
We implemented the FARS technique by using an electro-optic modulator (EOM$_{FARS}$, see Fig. \ref{setup}) which rapidly tunes the frequency of the modulator sideband over successive modes of the optical cavity (15~GHz tuning range).
With a standard \mbox{FS-CRDS} method, the frequency tuning takes much longer time than an individual \mbox{ring-down} decay (the decay time is of the order of 1~ms, while the laser tuning, cavity mode searching and PDH \cite{Drever1983} relocking takes from few seconds up to a minute) \cite{WCISLO201841}.
Therefore the implementation of the FARS technique allowed us to considerably reduce the experimental dead time related to laser tuning and relocking.
Furthermore, the ability to quickly jump over the cavity modes allowed us to completely reverse the sequence of a measurement cycle and instead of averaging the signal at every frequency point until reaching the Allan variance minimum, we scan the whole line spectrum in a sub-second time and average consecutive spectra, which considerably reduces the common experimental noise of the spectrum background.
We developed a system for cavity temperature stabilization that ensures 10~mK stability including temperature gradients \cite{Zaborowski_2017}, which reduces the slow drifts of a spectrum baseline and collisional line-shape parameters.

\section{\label{sec:level1}Analysis \& Discussion}
\begin{figure*}[th!]
\centering \includegraphics[width=0.97\textwidth]{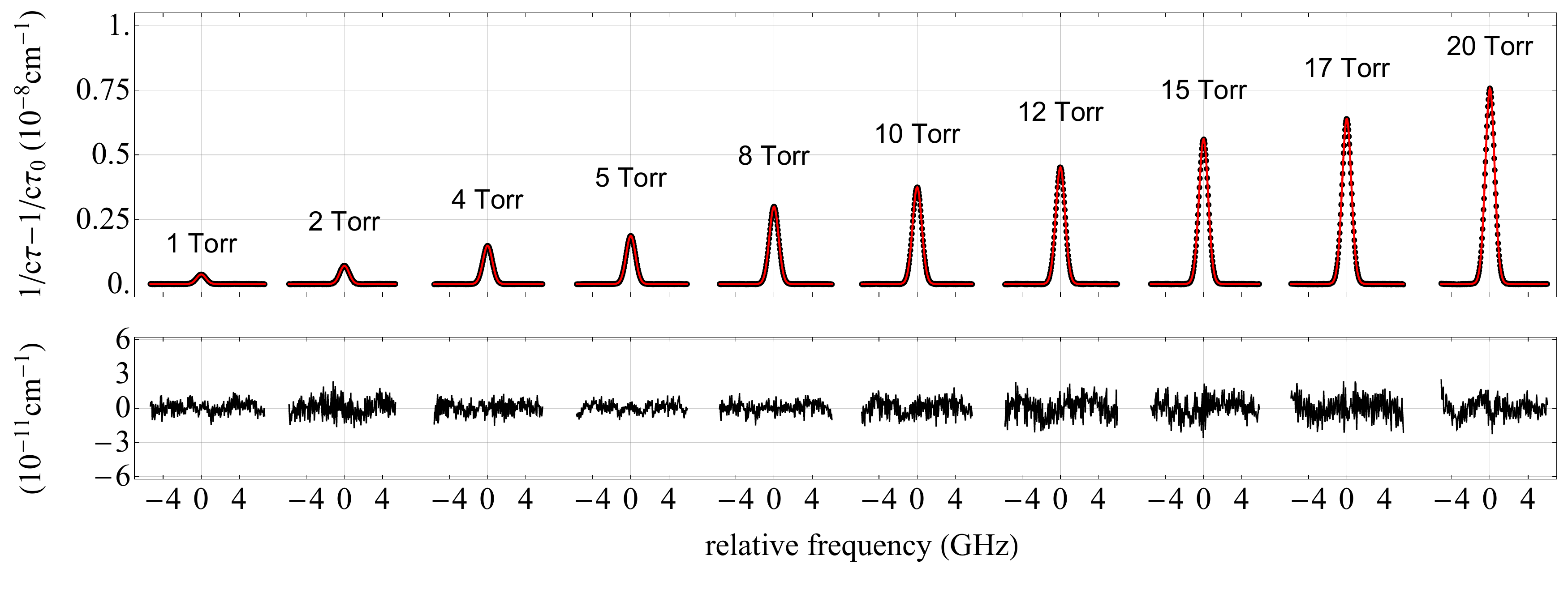}
\caption{(upper panel) 2--0 S(2) line of D$_2$ at 10~pressures from 1 to 20~Torr (black points are the experimental spectra and red lines are the fitted profiles).
(lower panel) The residuals obtained with SDBBP fit (standard deviation of residuals is 8$\times$10$^{-12}$~cm$^{-1}$ on average).}
\label{fig:residuals}
\end{figure*}

\begin{figure}[t!]
\includegraphics[width=0.97\columnwidth]{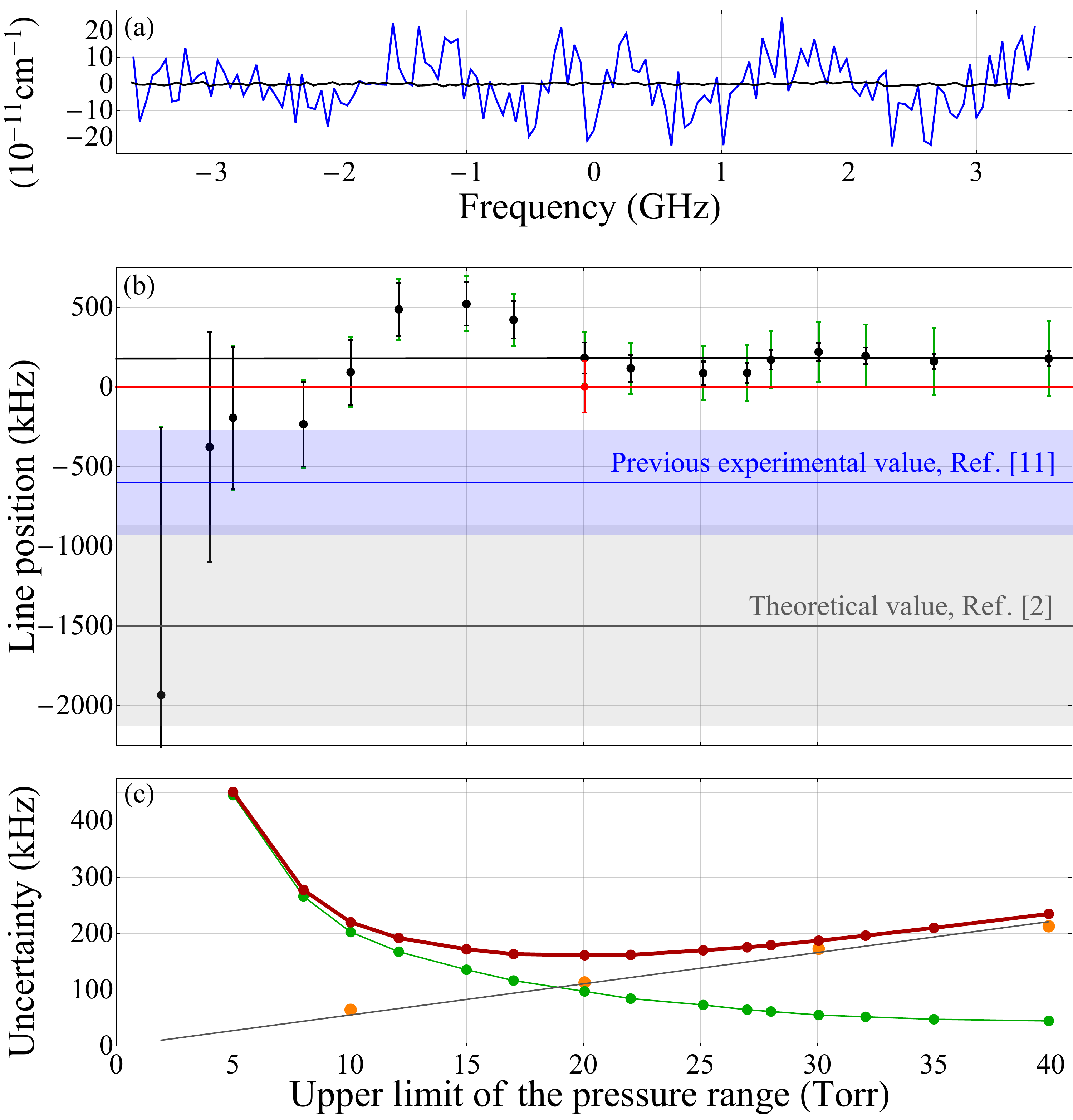}
\caption{
(panel~a) Residuals from the best fits of line-shape models to experimental spectra of the S(2) 2--0 line of D$_2$ for the case of: the present work for $4.9$~Torr (black line) and the previous work~\cite{WCISLO201841} for $984.4$~Torr (blue line).
The goal of this figure is not to compare the systematic structure of the residuals but the level of the noise before and after experimental improvements reported here.
The acquisition time is similar in both cases.
(panel~b) Comparison of our determination of the energy of the \mbox{S(2) 2--0} transition obtained with SDBBP fit (red point, 0~kHz value corresponds to 187~104~300.40(17)~MHz) with the best previous experimental result (blue line \cite{WCISLO201841}) and theoretical prediction (gray line \cite{PhysRevA.98.052506}); the blue and gray shadows are the corresponding uncertainties.
To show how the determined line position and its uncertainty depend on the choice of experimental conditions, we repeated the fitting analysis for different upper limits of the used pressure range, see black points.
Due to numerical expense of the SDBBP evaluation, we did this test with its approximated version, i.e., the beta corrected speed-dependent hard-collision profile ($\beta$SDHCP) \cite{WCISLO201675,KONEFAL2020106784}.
Although the line position determination with $\beta$SDHCP is systematically shifted from our ultimate value (by 179~kHz at 20~Torr upper limit) it properly shows the dependence of line position determination and its statistical uncertainty (black error bars) on the upper limit of the pressure range.
The green error bars are total uncertainties. 
(panel~c) Dependence of the total combined standard uncertainty of line position determination (red points) on the used pressure range of experimental data.
The gray line is the systematic part determined from 4~fits with perturbed values of the fixed line-shape parameters (orange points) and approximated with linear function.
The statistical part was determined as a standard uncertainty of the fitted line position (green points).}
\label{fig:together}
\end{figure}

The experimental setup developments allowed us to reach 37-fold lower noise-equivalent absorption level comparing to our previous experiment \cite{WCISLO201841}, see Fig.~\ref{fig:together}~(a), and, hence, move the experiment into the regime of much smaller pressures where the systematic uncertainties related to collisional effects are much smaller.
We collected the spectra of the S(2) 2--0 line in D$_2$ at 18~pressures from 1~to~40~Torr, see Fig.~\ref{fig:residuals}.
In this figure we show the spectra only for 10~pressures ranging from 1~to~20~Torr, which were used for our final determination of the line position.
It was shown in Ref.~\cite{PhysRevA.93.022501} that simple fits of the spectra with symmetric line profile and linear extrapolation of the line position to the zero-pressure limit results in a systematic error that comes from the fact that the effective line position does not scale linearly with pressure.
The reason is that the actual shapes of D$_2$ lines are asymmetric, mainly due to strong speed dependence of the collisional shift.
To reduce the influence of this asymmetry, in our analysis we used one of the most physically justified line-shape model describing the collisional effects, SDBBP, whose parameters are derived from \textit{ab~initio} quantum-scattering calculations and some of them were adjusted to the high-pressure spectra, see Ref.~\cite{WCISLO201841} for details.
The spectra for all the pressures were fitted simultaneously enforcing the same value of line position.
In contrast to ordinary multispectrum fit approach \cite{BENNER1995705,PINE2001180}, all the six collisional line-shape coefficients \cite{WCISLO201841} (i.e., the pressure broadening and shift, speed-dependence of the broadening and shift, and the real and imaginary parts of the complex Dicke parameter) and the Doppler broadening were fixed in experimental spectra fitting (this enforced also a proper linear scaling of the line-shape parameters with pressure).
Besides the common to all pressures line position, for each pressure we also fitted separate values of: the line area, linear baseline, and the amplitude and phase of a baseline etalon.
To estimate the influence of the fixed collisional line-shape parameters we repeated the fits with varied values of the collisional line-shape parameters by a conservative amount of 10\% \cite{WCISLO201841}.
In Figure~\ref{fig:together}, panels (b) and (c), we show how the line position determination and different sources of uncertainties depend on the pressure range taken in the analysis.
For every point in these plots we fitted all the spectra from the lowest pressure to the upper limit of the pressure range specified on the horizontal axis.
When only the low-pressure range is taken into account, the uncertainty is dominated by the statistical contribution and the uncertainty due to collisional effects is negligible.
In the opposite regime, in which all the pressures are included, the uncertainty is dominated by the contribution of the collisional perturbation of the line.
The smallest combined uncertainty is reached when the two dominating uncertainty sources (i.e., statistics and line-shape profile) are equal, which approximately corresponds to the upper limit of the pressure range equal to 20~Torr, see Fig.~\ref{fig:together}~(c).
In Table~\ref{tab:SCU}, we show the uncertainty budget.
Our ultimate determination of the D$_2$ 2--0 S(2) line frequency is 187~104~300.40(17)~MHz (wavenumber: 6241.127~667~0~(54)~cm$^{-1}$).
In Table~\ref{tab:results} and Fig.~\ref{fig:together}~(b), we show a comparison of our results with the recent theoretical value \cite{PhysRevA.98.052506} and previous experimental results \cite{MONDELAIN20165,WCISLO201841}.
The difference between theory and our results is 2.3$\sigma$.
The leading term in QED correction to the D$_2$ 2--0 S(2) line frequency is -0.033~167~(18)~cm$^{-1}$ \cite{PhysRevA.98.052506}, hence our experimental determination allows this correction to be tested at fifth meaningful digit.
Furthermore, our uncertainty is sufficiently small to test two other higher-order QED corrections and the finite nuclear size correction, see Table~$V$ in Ref.~\cite{PhysRevA.98.052506}.

\begin{table}[b!]
\caption{Standard uncertainty budget to experimental determination of the frequency of the S(2) 2--0 transition in D$_2$.
Middle and right column show uncertainties estimated for the fitted datasets of 1-20~Torr and 1-40~Torr, respectively.}
\renewcommand{\arraystretch}{1.25} \resizebox{\columnwidth}{!}{
\begin{tabular}{l d{3.2} d{3.2}}
\hline \hline Uncertainty contribution	& \mc{\makecell{u($\nu_0$) (kHz)\\1 to 20~Torr}}	& \mc{\makecell{u($\nu_0$) (kHz)\\1 to 40~Torr}} \\
\hline \hline
1. Line-shape profile					& 111			& 222\\
2. Statistics, 1$\sigma$				& 96			& 40\\
3. Instrumental systematic shift		& 47			& 47\\
4. Etalons								& 46			& 21\\
5. Temperature instability				& 5				& 5\\
6. Relativistic asymmetry				& 3				& 3\\
7. Optical frequency comb				& 1				& 1\\
8. Pressure gauge nonlinearity			& 1				& 1\\
\hline \textbf{Standard combined uncertainty}	& \textbf{161}	& \textbf{231} \\
\hline \hline 
\end{tabular}}
\label{tab:SCU}
\end{table}

\begin{table}[t]
\caption{A comparison of the present determination of the frequency of the D$_2$ 2--0 S(2) line with the previous experimental and theoretical values.}
\begin{center}
\begin{tabular}{l d{7.11}}
\hline
\hline
	& \mc{Line position (cm$^{-1}$)}\\
\hline
1. \textit{Ab~initio} theor.~\cite{PhysRevA.98.052506}						& 6241.127$ $617(21)\\
\makecell{2. Previous experiment,\\combined Refs.~\cite{WCISLO201841,MONDELAIN20165}} 	& 6241.127$ $647(11)\\
3. This work																& 6241.127$ $667$ $0(54)\\
\hline
\hline
\end{tabular}
\end{center}
\label{tab:results}
\end{table}

\section{\label{sec:level1}Conclusions}
We demonstrated an accurate measurement of the frequency of the weak S(2)~2--0 line in deuterium.
We reached the accuracy of 161~kHz by merging the FARS technique with a ultra-high finesse cavity.
The reported here value of the line position differs from previous experimental determination~\cite{WCISLO201841} by 600~kHz (1.6$\sigma$) and from theoretical value~\cite{PhysRevA.98.052506} by 1500~kHz (2.3$\sigma$).

\section{\label{sec:level1}Acknowledgements}
NS, HJ and PW contribution is supported by the National Science Centre in Poland through project no. 2018/31/B/ST2/00720.
The project is supported by the French-Polish PHC Polonium program (project 42769ZK for the French part).
The project is co-financed by the Polish National Agency for Academic Exchange under the PHC Polonium program (dec. PPN/X/PS/318/2018).
MZ and SW contribution is supported by the National Science Center, Poland, project no. 2017/26/D/ST2/00371.
AC, PM and DL contribution is supported by the National Science Center, Poland, project no. 2015/18/E/ST2/00585.
GK contribution is supported by NCN projects 2016/21/N/ST2/00334, 2017/24/T/ST2/00242.
MS and KS contribution is supported by the \textit{A next-generation worldwide quantum sensor network with optical atomic clocks} project carried out within the TEAM IV Programme of the Foundation for Polish Science cofinanced by the European Union under the European Regional Development Fund.
The research is a part of the program of the National Laboratory FAMO (KL FAMO) in Toru\'{n}, Poland, and is supported by a subsidy from the Polish Ministry of Science and Higher Education.

\section{\label{sec:level1}Disclosures}
\noindent\textbf{Disclosures.} The authors declare no conflicts of interest.

\bibliography{references}

\end{document}